\documentclass[9pt,twocolumn,twoside]{pnas-new}
\usepackage{amsmath}
\usepackage{amssymb}
\usepackage{graphicx}
\usepackage{xcolor}
\usepackage{hyperref}
\hypersetup{
    colorlinks = true,
    linkbordercolor = {white},
    allcolors = {blue}
}

\templatetype{pnasresearcharticle} 

\title{A Direct Two-Dimensional Pressure Formulation in Molecular Dynamics}

\author[a,1]{Sumith Yesudasan}
\author[b]{Sibi Chacko} 

\affil[a]{School of Chemical, Materials\\ and Biomedical Engineering\\
	University of Georgia\\
	Athens, Georgia 30606, USA}
\affil[b]{School of Engineering and\\Physical Science\\
    Heriot Watt University, Dubai Campus\\
    Dubai, UAE 294345}

\leadauthor{Yesudasan} 

\significancestatement{This paper presents an accurate mathematical method to estimate the pressure in an atomic system on a two dimensional plane. This method can replace the traditional three dimensional estimation of pressure and then averaging to a 2D grid and hence can save computational power.}

\authorcontributions{All authors equally contributed}
\authordeclaration{Authors have no conflicts of interests to declare}
\correspondingauthor{\textsuperscript{1}To whom correspondence should be addressed. E-mail: sumith.yd@uga.edu}

\keywords{Molecular dynamics $|$ Local pressure $|$ Multiscale coupling} 

\begin{abstract}
Coupling of statistical properties from atomistic simulations to continuum is essential to model many multi-scale phenomena. Often, the system under consideration will be homogeneous in two-dimensions (2-D). But due to the existing coupling methods, the property estimation takes place in three-dimensions (3-D) and then averaged to 2-D, which is computationally expensive due to the 3-D convolutions. A direct 2-D pressure or stress estimation model is lacking in literature. In this work, we develop a direct 2-D pressure field estimation method which is much faster than 3-D methods without losing accuracy. The method is validated with MD simulations on two systems: a liquid film and a cylindrical drop of argon suspended in surrounding vapor. This formulation will enable the study of 2-D fundamental phenomena like passive liquid flows in microlayer, as well as facilitate the coupling of atomistic and continuum simulations with reduced computational cost.
\end{abstract}

\dates{This manuscript was compiled on \today}

\begin{document}

\maketitle
\thispagestyle{firststyle}
\ifthenelse{\boolean{shortarticle}}{\ifthenelse{\boolean{singlecolumn}}{\abscontentformatted}{\abscontent}}{}

\dropcap{M}ulti-scale coupling of atomistic and continuum simulations is of significant importance in the areas of heat transfer, fracture mechanics and bioengineering \cite{horstemeyer2009multiscale}. For example, in the study of bio-membrane bending, it is necessary to understand the local variations of pressure and stress which are inaccessible through experiments \cite{lindahl2000spatial, tieleman1997computer, berger1997molecular, feller1999constant}. These computations typically map properties determined from atomistic simulations onto grid points in continuum simulations \cite{2steinhauser2008computational}, which is used for estimating inter facial energies, surface tension, pressure gradients in fluid simulations and lipid bilayer mechanics. Many simulated systems have inhomogeneity only in two dimensions (2-D) such as defect nucleation in bulk and 2-D crystals, bio molecular assemblies such as lipid bilayers and membrane proteins, as well as thin film evaporation and heat transfer \cite{maroo2015surface,maroo2016origin, daisy2016molecular}, and thus only require 2-D pressure distribution. However, current literature on local pressure estimation is based on 3-D \cite{3ollila20093d} or 1-D \cite{4sonne2005methodological,5lindahl2000spatial,6heyes2011equivalence} pressure estimation. The 2-D pressure distribution is obtained by averaging over the 3-D pressure data, and is extremely computationally expensive \cite{7vanegas2014importance} as it involves a 3-D convolution. A generalized method for 3-D stress calculations which included temporal averaging weight functions was derived by Yang \cite{8yang2012generalized}. Recently, Vanegas \cite{7vanegas2014importance} and Sanchez et al. \cite{9torres2015examining} applied the modified Hardy versions of IK stress to lipid bilayers, coiled coil protein and graphene sheet to determine continuum level properties from atomistic simulations.  Further, there exist a few Irving-Kirkwood versions \cite{10lee2012pressure,11weng2000molecular,12hardy1982formulas} of 1-D pressure calculations for 1-D inhomogeneous system. However, to the best of our knowledge, no methods are present for a direct 2-D pressure estimation. 

This work presents a 2-D pressure estimation algorithm based on Hardy's stress method, which is validated by performing molecular simulations of a suspended liquid film and a cylindrical drop and comparing the results with experimental data and classical Young-Laplace equation, respectively. This can be very useful not only in atomic scale systems but also in mesoscale dynamics with continuum coupling \cite{yesudasan2018molecular,yesudasan2018fibrin,yesudasan2018molecular2,yesudasan2018coarse}.

Historically, the atomic level virial stresses from statistical analysis were first derived by Irving and Kirkwood \cite{13irving1950statistical}, now generally referred to as IK method. The need for large ensemble averaging due to the delta function in IK method was circumvented by Hardy in his classical paper \cite{12hardy1982formulas,14hardy2004two} by introducing a smoothing function and a bond function. The virial stress has two components, a kinetic component and a force component. There existed an ambiguity among researchers about the equivalence of virial stress with Cauchy stress. The ambiguity is thoroughly discussed in Zhou's paper \cite{15zhou2003new} which claims that Cauchy stress is not equivalent to virial stress, but is equivalent only to the force component of virial stress. Based on this finding, researchers \cite{16zimmerman2004calculation, 17gall2004strength, 18Buehler2006Dynamical, 19gates2005computational} performed a number of molecular studies. Zimmerman \cite{16zimmerman2004calculation} showed that, for crystals, Hardy's stress formulation gave more accurate results than simple local virial averages. A comparative study of different versions of local virial stress was studied by Murdoch \cite{20murdoch2007critique}. In contrast to Zhou's work \cite{15zhou2003new}, Subramaniyan \cite{21subramaniyan2008continuum} found that virial stress is indeed the Cauchy stress using specific examples. There were other works \cite{22admal2010unified,23zimmerman2010material} which tried to develop the appropriate relation of virial stress and continuum level stresses. 

All these studies are performed in 3-D domain and later averaged to 2-D. A consistent direct 2-D formulation of local pressure is missing in the literature. In this paper we will derive 3-D, 2-D and 1-D versions of local pressure estimation. This will be used to estimate pressure, density and temperature of certain case studies and will be validated. Our work also supports the fact that while converting virial stress to a continuum level property, both kinetic component and force component of virial stress should be considered.
\subsection*{Local pressure estimation}
The 3-D pressure from molecular interactions is estimated classically by IK method \cite{13irving1950statistical} through the expression shown in Eq. \ref{eq1}. Here, the first term represents the kinetic energy contribution and second represents the virial contribution. 
\begin{equation}
P(r_p) = P_K(r_p) + P_V(r_p)
\label{eq1}
\end{equation}
Kinetic contribution is,
\begin{equation}
P_K(r_p) = \sum^N_{i=1} m_i v_i \circledast v_i \delta(r_i-r_p)
\label{eq2}
\end{equation}
Virial contribution is
\begin{equation}
P_V(r_p) = \sum_{i=1}^{N-1} \sum_{j=i+1}^{N} r_{ij} \circledast F_{ij} \delta(r_i-r_j)\delta(r_i-r_p)
\label{eq3}
\end{equation}
Here $P$ is the pressure, $m$ is mass of $i^{th}$ atom, $v$ is velocity, $r_i$ and $r_j$ are the position vectors of $i^{th}$ and $j^{th}$ atoms respectively, $N$ is number of atoms, $r_p$ is the position vector of $p^{th}$ grid point, $r_{ij} = r_i -r_j$, $F_{ij}$ is the force, and $\delta$ is the Dirac delta function \cite{dirac1981principles}. Though this expression is theoretically correct, practically it needs infinite sampling, which makes it less appealing for finite computer simulations. Specifically, for molecular dynamics simulations this is computationally expensive due to its convolution nature. To evade this situation, Hardy introduced \cite{12hardy1982formulas,14hardy2004two} interpolation functions to distribute the kinetic contribution and a bond function to distribute the virial contribution to the local grid points. This resulted in the modified expression for pressure as
\begin{equation}
P(r_p) = \sum_{i=1}^{N} m_i v_i \circledast v_i w(r_i - r_p) + \sum_{i=1}^{N-1} \sum_{j=i+1}^{N} r_{ij} \circledast F_{ij} B_{ij}(r_p)
\label{eq4}
\end{equation}
Here, $w$ is the weight function (interpolation function) and $B$ is the bond function and defined as
\begin{equation}
B_{ij}(r_p) = \int_{0}^{1} w(\lambda r_{ij} + r_i - r_p)~ d\lambda
\label{eq5}
\end{equation}
A weight function has to be normalized and should follow
\begin{equation}
\int_{R^3} w(r)dr^3 = 1
\label{eq6}
\end{equation}

\subsection*{3-D pressure formulation}
For a 3-D system, if the distribution is assumed to be spherically symmetric, then 
\begin{equation}
\int_{R^3} w(r)dr^3 = \int_0^\infty \hat{w}(r) 4\pi r^2 dr = 1
\label{eq7}
\end{equation}
If the spread of the function is limited to a certain spread radius $r_s$ then the equation becomes
\begin{equation}
\int_{R^3} w(r)dr^3 = \int_0^{r_s} \hat{w}(r) 4\pi r^2 dr = 1
\label{eq8}
\end{equation}
Here, $\hat{w}(r)$ is the weight function and used by researchers \cite{8yang2012generalized,22admal2010unified} for 3-D grid, is given as:
\begin{equation}
\hat{w}(r) = C_1 [1 - 3r^2/r_s^2 + 2r^3/r_s^3]
\label{eq9}
\end{equation}
here, $C_1$ is the normalization constant.
For 3-D systems, 
\begin{equation}
\int_{R^3} w(r)dr^3 = \int_0^{r_s} 4\pi r^2 C_1 [1 - 3r^2/r_s^2 + 2r^3/r_s^3] dr = 1
\label{eq10}
\end{equation}
the constant of integration takes the form $C_1 = 15 / 4 \pi r_s^3$ and $r(x,y,z)$ is a function in three coordinates.

\subsection*{2-D pressure formulation}
\begin{figure}[!httbp]
\includegraphics[width=1\linewidth]{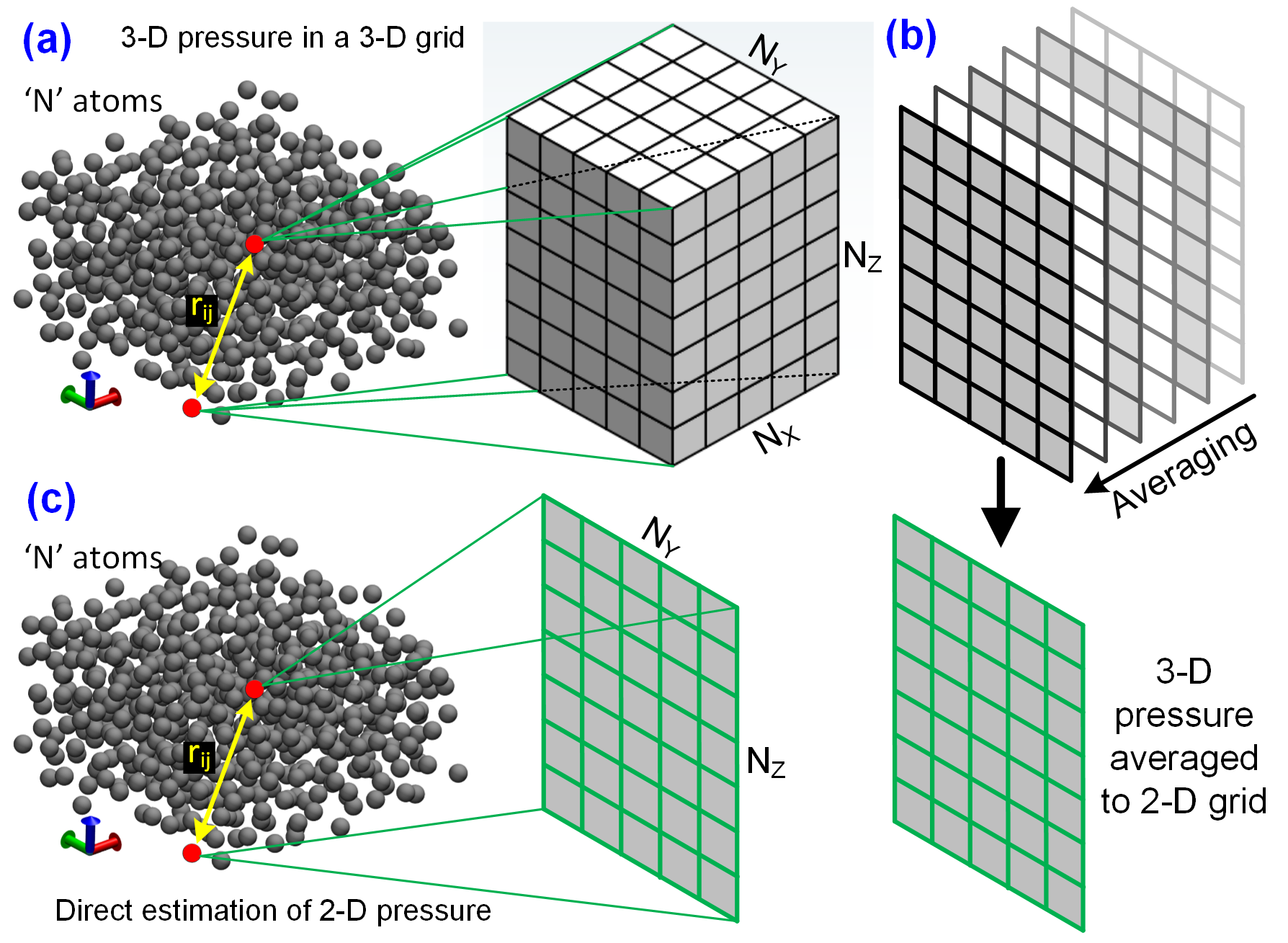}
\caption{Local pressure estimation in 3-D and 2-D grids. (a) Schematic of pressure estimation in a 3-D grid from a molecular system. (b) Estimation of pressure in a 2-D grid by averaging the 3-D grid data (This is the traditional approach). (c) Direct estimation of pressure in 2-D grids from the MD simulation system (this work).}
\label{figure1}       
\end{figure}

In this section we will explain the formulation of 2-D local pressure method by reformulating the 3-D weight function which will significantly reduce the computational cost without losing any desired details in the results. Typically, a 3-D local pressure method requires $N^2 \times N_X \times N_Y \times N_Z \times N_B$ operations ($N$ is the number of atoms; $N_X$, $N_Y$ and $N_Z$ are the number of grid points along x, y and z-directions respectively; $N_B$ is the number of discrete points for bond function integration). Here, the first term $N^2$ is the cost of inter-atomic pair potential force determination, which can be reduced to $O(N)$ using cell list algorithms \cite{26welling2011efficiency}. This will make the 3-D pressure estimation cost as $N \times N_X \times N_Y \times N_Z \times N_B$ as shown in the Fig. \ref{figure1}a.  

\begin{figure}[!httbp]
\includegraphics[width=1\linewidth]{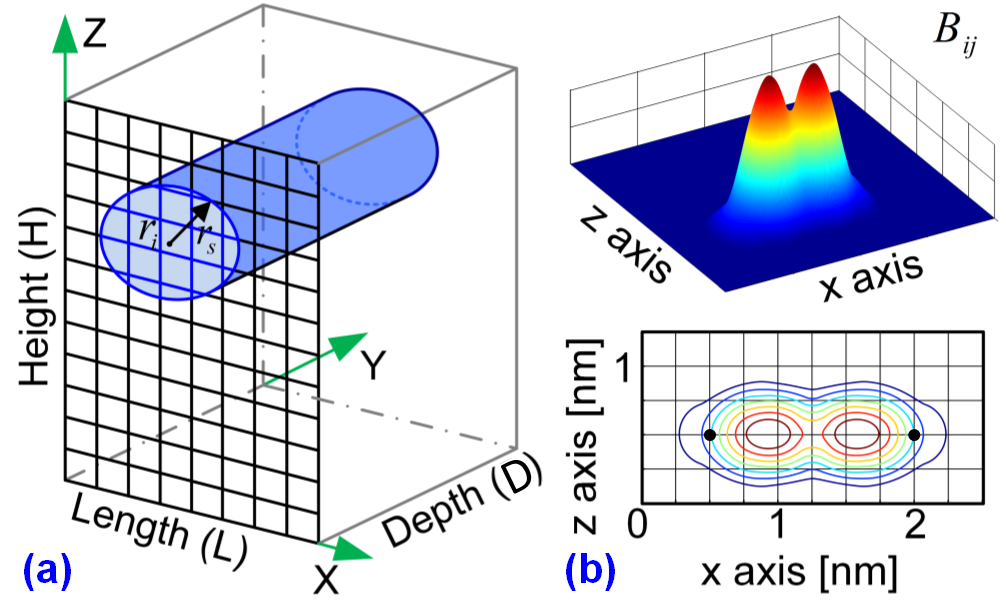}
\caption{Weight and bond function developed for 2-D pressure formulation. (a) Cylindrical volume of influence associated with an atom located at , where   is the spread radius, L, D, H are length, depth and height respectively. (b) Visualization of bond function for two atoms separated at a distance of 1.5 nm. The gradient image (upper) shows variation along surface and the contour plot of the same is shown below.}
\label{figure2}       
\end{figure}

While estimating the pressure in a 2-D grid, traditionally, the pressure in the 3-D grid is averaged to obtain it as seen in the Fig. \ref{figure1}b. This expensive step will become unnecessary if we can directly estimate the pressure in 2-D grids as shown in Fig. \ref{figure1}c. Though it looks like a trivial case, the results are very promising by reducing the computational effort to $N \times N_X \times N_Z \times N_B$. In this work, we propose that while extending the pressure estimation theory to a 2-D grid, the spherical distribution volume has to be changed to a cylindrical volume as shown in Fig. \ref{figure2}a. This is the case with most of the 2-D non-homogeneous systems. 

The thermodynamic property variations along the y-axis is considered unchanged over long period of time and hence the $r(x,z)$ depends only on $x$ and $z$. 
The resulting 2-D weight function will follow:
\begin{equation}
\int_{R^3} w(r)dr^3 = \int_0^{r_s} 2\pi r D C_1 [1 - 3r^2/r_s^2 + 2r^3/r_s^3] dr = 1
\label{eq11}
\end{equation}

Here, $D$ is the depth of the system along $Y$ (direction of homogeneity) as shown in Fig. \ref{figure2}a, $r_s$ is the \textit{spread} radius and the constant of integration is $C_1 = 10/3 \pi D r_s^2$. 

Figure \ref{figure2}b shows the variation of bond function for a pair of atoms in the case of 2-D system kept at $1.5~ nm$ apart. The isometric view shows the variation of magnitude of bond function for a spread radius of $0.5~ nm$. 

Thus, based on the new weight function, the density of the system ($\rho$) is defined as:
\begin{equation}
\rho(r_p) = \sum_{i=1}^{N} m_i \hat{w}(r_i - r_p)
\label{eq12}
\end{equation}
local number density at a grid point is
\begin{equation}
n(r_p) = \sum_{i=1}^{N} \hat{w}(r_i - r_p)
\label{eq13}
\end{equation}
and temperature as 
\begin{equation}
T(r_p) = \sum_{i=1}^{N} \frac{m_i v_i \circledast v_i}{3 n(r_p) k_B}  \hat{w}(r_i - r_p)
\label{eq14}
\end{equation}

The selection of our interpolation function is arbitrary to demonstrate the 2-D formulation and instead any of the popular functions can be used. With that in mind, we have formulated 2-D forms for some selected functions, along with their 3-D functions are shown below.

\noindent Quadratic:
\begin{align}
&\hat{w}_{3D}(r) = \frac{15(1-r^2/r_s^2)}{8 \pi r_s^3}\\
&\hat{w}_{2D}(r) = \frac{2(1-r^2/r_s^2)}{D \pi r_s^2}
\label{eq15}
\end{align}

\noindent Exponential:
\begin{align}
&\hat{w}_{3D}(r) = \frac{2.2671}{ r_s^3} exp(\frac{r_s^2}{r^2-r_s^2})\\ 
&\hat{w}_{2D}(r) = \frac{2.1435}{D r_s^2} exp(\frac{r_s^2}{r^2-r_s^2})
\label{eq16}
\end{align}

\noindent Trignometric:
\begin{align}
&\hat{w}_{3D}(x,y,z) = \frac{1}{ 8 r_s^3} (1+cos(\frac{\pi x}{r_s}))(1+cos(\frac{\pi y}{r_s}))(1+cos(\frac{\pi z}{r_s}))\\ 
&\hat{w}_{2D}(x,z) = \frac{1}{ 4D r_s^2} (1+cos(\frac{\pi x}{r_s}))(1+cos(\frac{\pi z}{r_s}))
\label{eq17}
\end{align}

For grid dependent and finite support weight functions like B-splines, a rectangular prism volume could be used instead of cylindrical volume.

\subsection*{1-D pressure formulation}
For completeness, we have also derived the 1-D variation of pressure and density which is very suitable for 1-D inhomogeneous systems like pressure in thin films, lipid bilayers etc. The $r(z)$ will now depend only on the z-axis and the x and y axis variations are assumed to be negligible over time. 
\begin{equation}
\int_{R^3} w(r)dr^3 = \int_0^{r_s} 2 L D C_1 [1 - 3r^2/r_s^2 + 2r^3/r_s^3] dr = 1
\label{eq18}
\end{equation}

This will give the integration constant as $C_1 = 1/LDr_s$. This is also consistent with the derivation of Hardy stress \cite{12hardy1982formulas} and will be shown with example results in the next section.

\subsection*{Results and discussion}

\begin{figure}[!ht]
\includegraphics[width=1\linewidth]{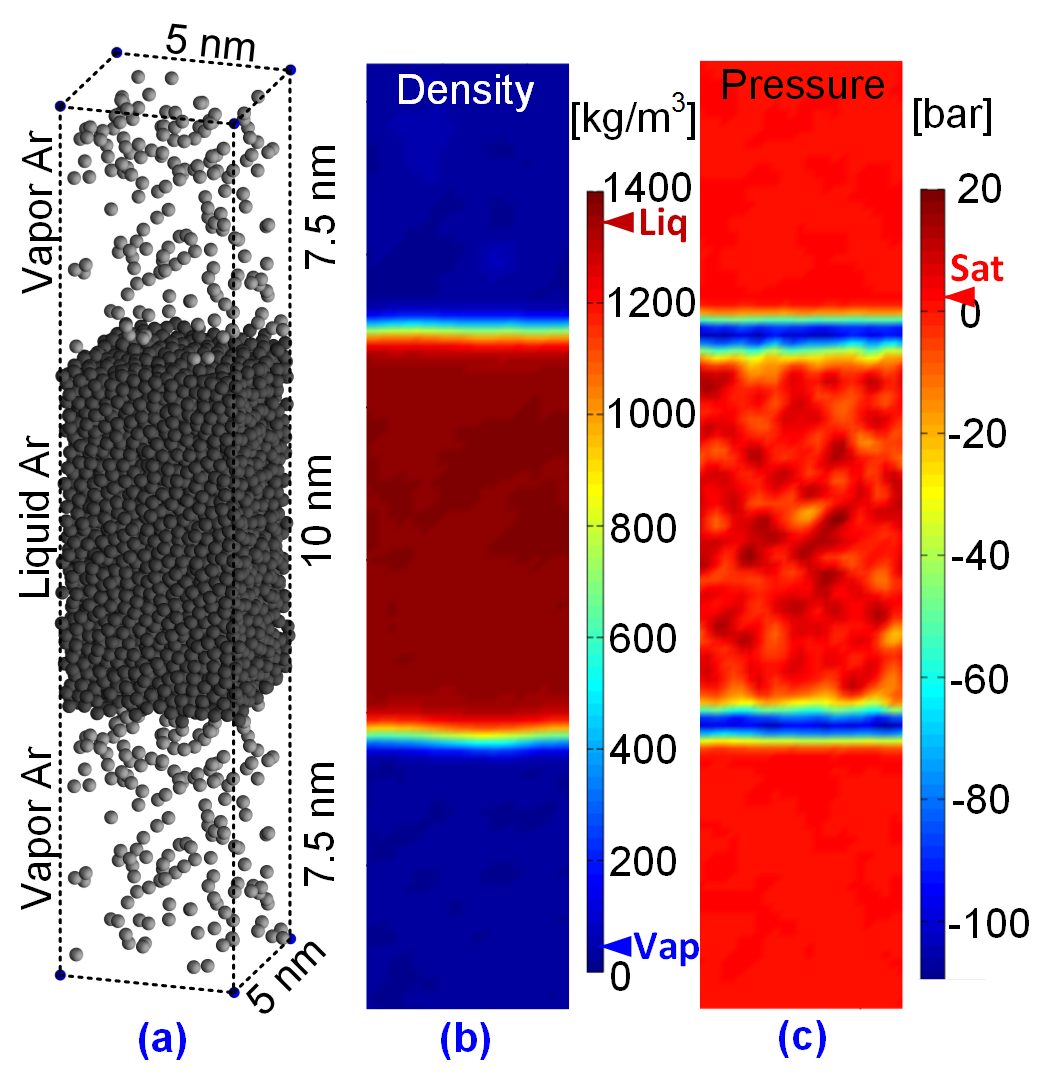}
\caption{Two-dimensional density and pressure profile in argon multiphase system using the new 2-D formulation. (a) A $10 ~nm$ thick argon film suspended with $7.5 ~nm$ thick vapor on both sides along the z-direction. Two-dimensional (b) density and (c) pressure distribution obtained for the system equilibrated at $90~K$. The saturation density (NIST data) corresponding to liquid (Liq) and vapor (Vap) are marked in the density plot colorbar, while the saturation pressure (Sat) corresponding to the saturated fluid at $90~ K$ (NIST data) is marked in the pressure plot colorbar showing good agreement with the simulation results.}
\label{figure3}       
\end{figure}

In order to demonstrate and validate the new 2-D pressure formulation, we apply it to study the pressure, surface tension and density variations of argon liquid films suspended in argon vapor using MD simulations. In our chosen example (argon liquid film suspended in vapor as shown in Fig. \ref{figure3}a) and also for lipid bilayer \cite{7vanegas2014importance}, the inhomogeneity is in two dimensions (say, $X$ and $Z$ axes) and there is no bulk density variation along the third dimension ($Y$ axis) over ensemble average. 

\begin{figure*}[!ht]
\includegraphics[width=1\linewidth]{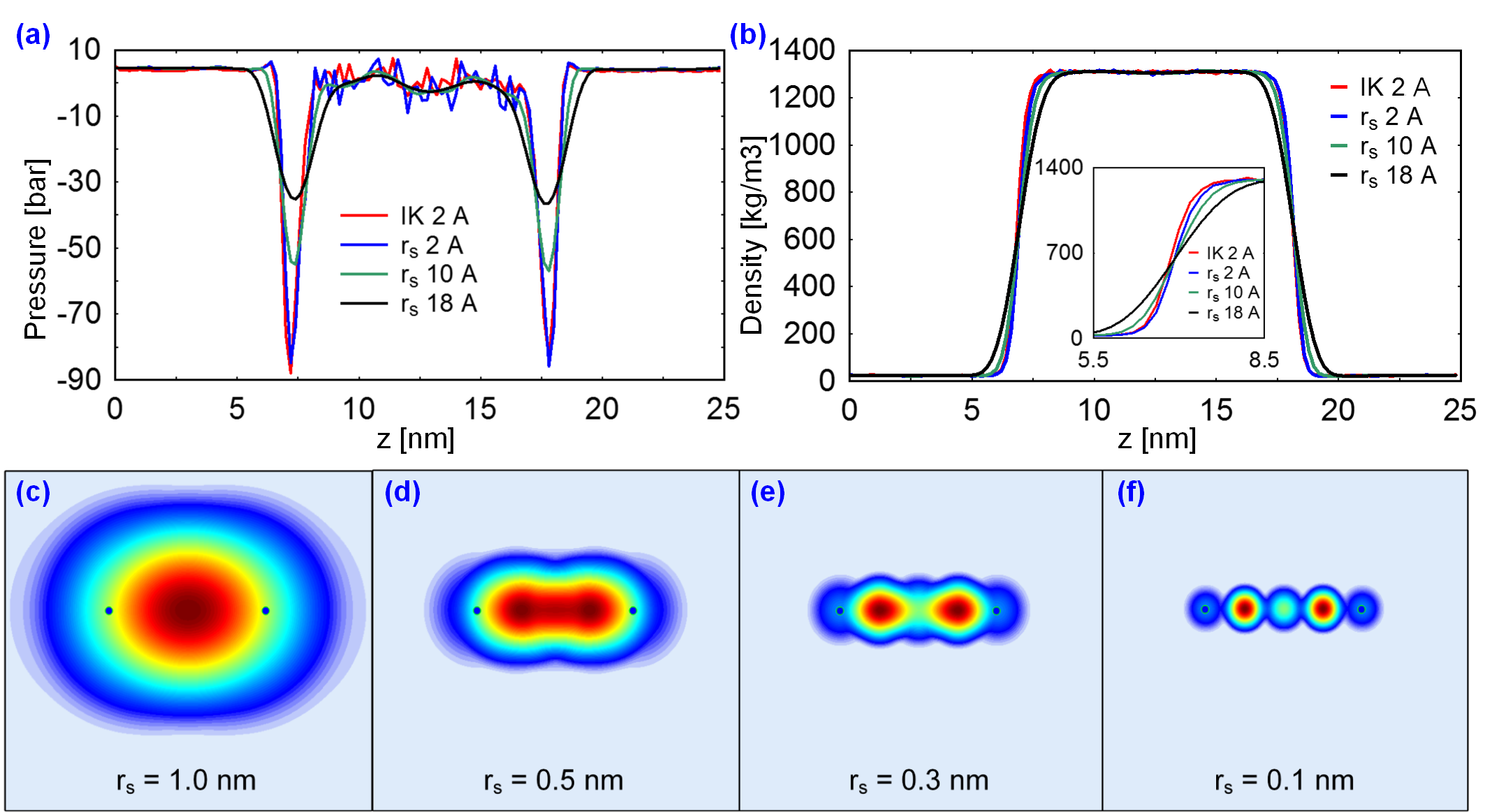}
\caption{Sensitivity study of spread radius $r_s$ on bond function, pressure and density. (a) Pressure variation across the argon film for different values of $r_s$. "IK 2 A" is the case study using the established Irving-Kirkwood's modified 1-D implementation \cite{11weng2000molecular} for comparison. The "IK 2 A" and the new 2-D formation based profiles show good agreement when the volume of smearing became comparable. (b) Density variation across the film for different values of  which confirms that the overall system bulk properties is not affected by the spread radius. (c-f) Contour plots of bond function with $r_s$ ranging from $1 ~nm$, $0.5~ nm$, $0.3~ nm$ and $0.1 ~nm$ for two atoms kept $1.2~ nm$ apart in a $3 ~nm \times 3 ~nm$ domain. The plots visually show how the bond function controls the spreading of the pressure and density across the grids for different spread radii.}
\label{figure4}       
\end{figure*}

The computational domain is shown in Fig. \ref{figure3}a. The argon liquid film is $10~ nm$ thick with $7.5~ nm$ thick argon vapor on either side along the z-direction. The X-Y cross section size is $5~ nm \times 5~ nm$. Periodic boundary conditions are applied in all directions. The vapor and liquid domains in this molecular system are first equilibrated separately \cite{11weng2000molecular} for $1000 ~ps$ in order to get a stable suspended film and are then brought them together. The system is then equilibrated for $1000 ~ps$ followed by production run for another $1000 ~ps$ on which statistical analysis is performed. The modified Stoddard-Ford LJ potential \cite{24stoddard1973numerical} is used with argon – argon LJ parameters as $\sigma_{Ar-Ar} = 0.34 ~nm$  and $\epsilon_{Ar-Ar} = 1.005841 ~kJ/mol$. The time step of velocity verlet integration is $5~ fs$ and the thermostat to keep temperature constant is chosen as velocity scaling algorithm. MD simulations for different temperatures, spread radius and cutoff radius were performed. A validated, self-written C++ molecular dynamics code is used for all simulations \cite{daisy2016molecular}.

It is found that the thermodynamic properties like pressure of argon is best captured by using a cutoff radius of $5 \sigma$ or greater \cite{weng2000molecular}. This corresponds to $1.8 ~nm$ for argon and we have used the same for all the simulations presented in this work. In the literature, it is common to consider the cutoff radius $r_c$ of MD simulations and spread radius $r_s$ of local pressure calculation as the same. However, considering same cutoff and spread radius will lead to increased number of grid point influence, increasing the computational cost and also limits the finer local details. Hence, in this work, the dependency between spread radius and cutoff radius is removed and considered them as separate entities, which enables us to retain the accuracy of the simulation without introducing any artifacts by choosing a higher cutoff radius. Therefore, the spread radius can be adjusted to capture the localized effects as desired.

Using the developed 2-D formulation, the temporally averaged 2-D contours of density and pressure at $90 ~K$ are estimated and shown in Figs. \ref{figure3}b and \ref{figure3}c. The density and pressure results are compared with the saturation properties from NIST thermodynamic properties database \cite{25lemmon2005thermophysical} and found to be in very good agreement, which highlights the accuracy of the pressure and density calculation in the new formulation. 

\begin{figure*}
\includegraphics[width=1\linewidth]{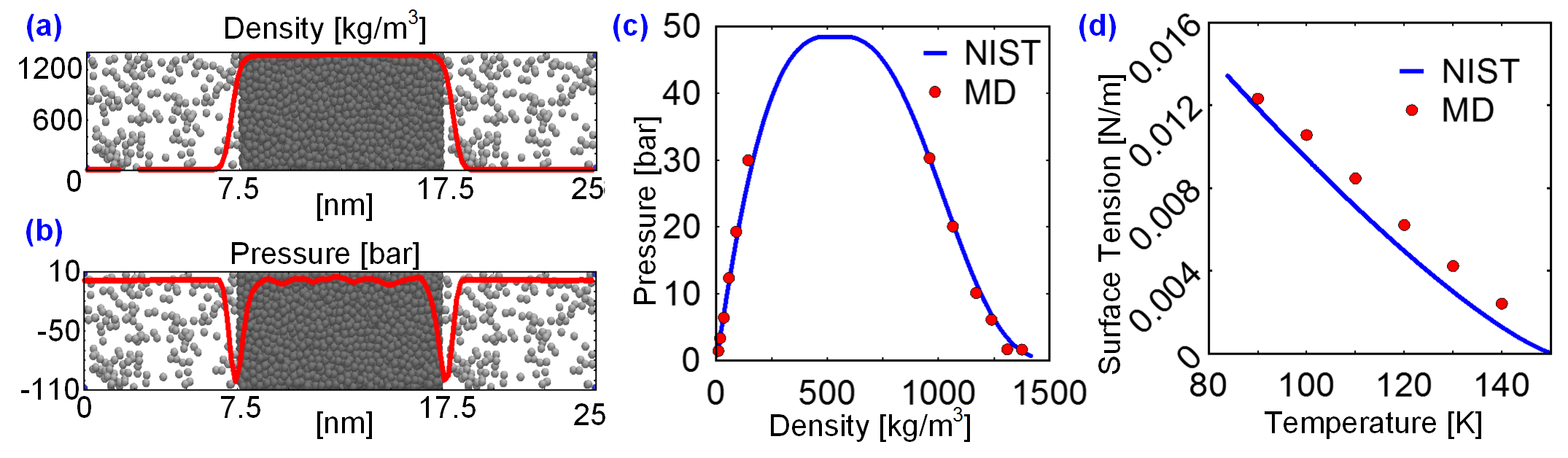}
\caption{Comparison of MD simulation results with the standard thermodynamic physical data from NIST \cite{25lemmon2005thermophysical}. (a) 1-D density profile and (b) 1-D pressure profile, deduced from the new 2-D formulation method, plotted over the molecular simulation of argon film. The interface locations capture the expected change in pressure and density. Comparison of MD simulation results and thermodynamic data for (c) pressure vs. density, and (d) surface tension vs. temperature showing excellent agreement. Pressure is estimated by temporal and spatial averaging of vapor and liquid regions separately.}
\label{figure5}       
\end{figure*}

The sensitivity of spread radius on pressure and density results is studied using the system shown in Fig. \ref{figure3}a by varying the spread radius to $0.2 ~nm$, $1~ nm$ and $1.8 ~nm$ and estimating the 2-D properties of pressure and density. The 2-D values are then averaged along the  axis to obtain a 1-D pressure and 1-D density profile varying along the z-direction as shown in Figs. \ref{figure4}a and \ref{figure4}b respectively. Alongside, the pressure and density calculation based on the already-established 1-D IK method \cite{11weng2000molecular} with a slab thickness of 0.2 nm are also plotted. The results in Figs. \ref{figure4}a and \ref{figure4}b show that density and pressure smoothen and spreads to a larger area as the spread radius is increased. Also, when the spread radius is small and comparable to the slab thickness of IK method, both density and pressure matches very well. As expected, the bulk region (vapor only and liquid only) properties are found to be not sensitive to the spread radius since it primarily captures the local effects.

\begin{figure*}
\begin{center}

\noindent \includegraphics[width=.9\linewidth]{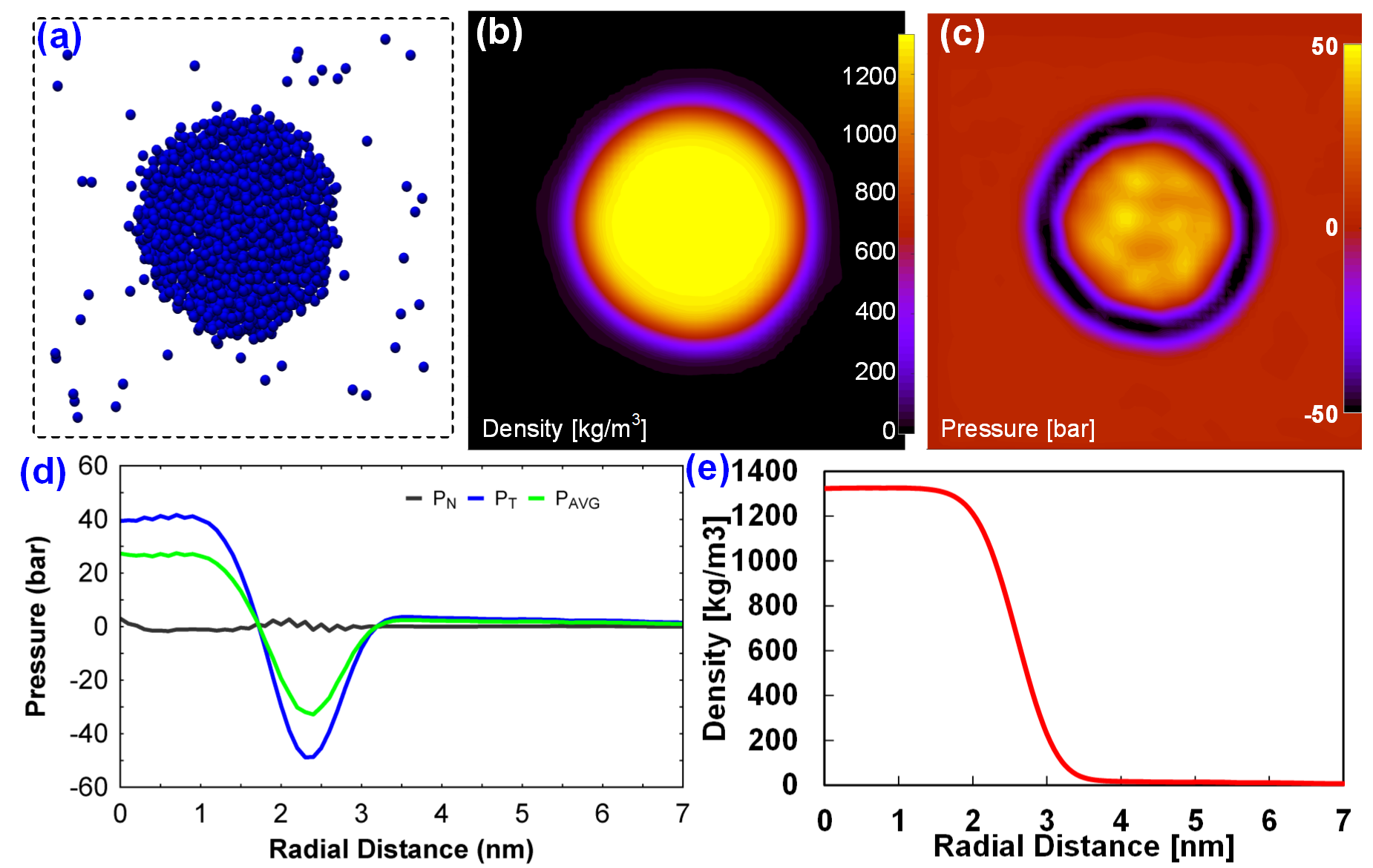}
\caption{Laplace pressure study in a cylindrical liquid argon droplet. (a) Molecular model of cylindrical argon in a 3-D periodic box. (b) Density of the system after ensemble averaging using our 2-D method. (c) Pressure of the system ensemble averaged using our 2-D method. (d, e) Pressure and density variation in the cylindrical Argon system using cylindrical coordinate conversion.}
\label{figure6}       

\end{center}
\end{figure*}

Further, to understand the dependency of the bond function to the spread radius, the bond function for two atoms placed at $1.5 ~nm$ apart are plotted with varying spread radius of $1 ~nm$, $0.5 ~nm$, $0.3 ~nm$ and $0.1 ~nm$ (Figs. \ref{figure4}c-f). The resulting images show an important result: the spread radius determines the degree of sharpness required to capture the local features as desired. Further, as long as the integral of bond function is unity and conserved, it does not give erroneous values for surface tension, density or pressure. However, care should be taken while selecting the grid cell size for smearing as the results may be less accurate when the spread radius becomes comparable to grid size (although the resulting artifacts can possibly be alleviated using finite support weight functions like B-Splines, which however needs further investigation).

Next, we validate the 2-D pressure formulation by performing multiple simulations with varying temperature of the argon system ($90~ K$, $100 ~ K$, $110 ~ K$, $120 ~ K$, $130 ~ K$, and $140 ~ K$) and comparing the simulation results with the experimental thermodynamic properties of argon from NIST database \cite{25lemmon2005thermophysical}. The spread radius and cutoff radius are chosen as $0.5 ~ nm$ and $1.8 ~ nm$ respectively for these simulations. We would like re-emphasize the fact that spread radius does not alter any continuum level quantities and the choice of $0.5 ~ nm$ as the spread radius is merely arbitrary. Thermodynamic quantities of pressure, density and surface tension are estimated using the developed 2-D methodology. The 2-D results are averaged along the x-direction to obtain a 1-D pressure and 1-D density profile varying along the z-direction. A visualization of pressure and density variation along the height of the domain is shown in Fig. \ref{figure5}a and \ref{figure5}b which is consistent with previous argon film studies \cite{10lee2012pressure,11weng2000molecular}. The comparison of pressure vs. density and surface tension vs. temperature are plotted in Figs. \ref{figure5}c and \ref{figure5}d, respectively, and show very good agreement with the experimental data \cite{25lemmon2005thermophysical}. 

Since the above system is inhomogeneous only in one-dimension, we performed another validation on a curvilinear system which is inhomogeneous in two-dimensions. We estimate the pressure difference in a cylindrical droplet as shown in Fig. \ref{figure6}a, and compare the result with the classical Young-Laplace equation. The droplet is symmetric in the plane of the figure with a depth of $3 ~nm$ and has periodic boundary conditions in all directions with sides of $11 ~nm$ each. The droplet is equilibrated for $1000 ~ps$ and then production runs are done for another $2000 ~ps$. The pressure and density is estimated at every 20 steps and averaged using the method introduced in this work. However, during the course of the simulation, the center of the droplet may vary around the original location. In order to avoid a skewed averaging, center of mass of every data set is found and readjusted to the center of the domain before averaging. The resulting ensemble averaged density and pressure is shown in Figs. \ref{figure6}b and \ref{figure6}c respectively. The variation of the pressure and density from the center of the droplet towards outside is shown in Figs. \ref{figure6}d and \ref{figure6}e. 

The excess pressure inside the drop is given by the classical Young-Laplace equation:
\begin{equation}
P_{in}-P_{out}=\frac{2\gamma}{R}
\label{eq19}
\end{equation}

where $P_{out}$ and $P_{in}$ are the outside and inside pressures of the drop, $\gamma$ is the surface tension, and $R$ is the radius of the drop. The radius $R$ is estimated by identifying the interface using our interface detection algorithm \cite{daisy2017robust, yd2015new, yesudasandaisy_2015}. All parameters in Eq. (\ref{eq19}) are estimated independently from the MD simulations. For the system simulated, we obtain  from the density profile, and the surface tension is estimated. In order to estimate the radial variation of the properties like normal pressure, density, tangential pressure and surface tension, we used the 2-D rotation matrix in combination with B-spline interpolation polynomials. The left hand side of Eq. (\ref{eq19}) results in a value of $3.3~ MPa$, while the right hand side results in $2.5~ MPa$, and thus, is in good agreement with the Young Laplace equation. We expect the agreement to improve further for larger drop sizes (however, with added computational cost). These simulations confirm the validity and accuracy of the new 2-D formulation method developed and presented in this work.

\subsection*{Conclusions}

In conclusion, a grid based method for two-dimensional estimation of pressure and density was developed and validated. The methodology was applied to a suspended argon liquid film in argon vapor with varying temperatures, and results were in very good agreement NIST experimental database values. The method was also applied to the classical problem of pressure difference calculation in a cylindrical drop and the results were found to be in good agreement with the Young Laplace equation. Further, the dependency between spread radius and cutoff radius was disconnected which allows for high accuracy of the simulation by choosing a higher cutoff radius without introducing any artifacts. The spread radius can be adjusted to capture the localized effects in the system as desired. The developed method will be significantly faster (computationally) than the existing 3-D grid method, and can be very useful in determining stresses occurring in lipid bilayers and other systems where inhomogeneity exists only in two of the three dimensions. This work also supports the fact that for the conversion of virial stress to a continuum level property, both kinetic component and force component of virial stress should be considered.

\textbf{Nomenclature}

\begin{tabular}{ l l }
1-D & One-dimensional\\
2-D& Two-dimensional\\
3-D & Three-dimensional\\
$B_{ij}$& Bond function between $i^{th}$ and $j^{th}$ atoms\\
$C_1$& Constant of integration\\
D & Depth\\
$F_{ij}$ & Force between $i^{th}$ and $j^{th}$ atoms\\
H & Height\\
IK & Irving-Kirkwood\\
L & Length\\
LJ & Lennard Jones\\
MD & Molecular Dynamics\\
N& Number of atoms\\
P & Pressure\\
$P_{in}$ & Pressure inside cylindrical drop\\
$P_K$ & Kinetic component of pressure\\
$P_{out}$ & Pressure outside cylindrical drop\\
$P_V$ & Virial component of pressure\\
R& Radius of cylindrical drop\\
T& Temperature\\
$k_B$& Boltzmann constant\\
$kJ$& kilo Joules\\
$m_i$ & Mass of $i^{th}$ atom\\
n & Number density of atoms in $p^{th}$ grid point\\
nm& nano meter\\
ps& pico second\\
$r_i$ & Position coordinate of $i^{th}$ atom\\
$r_p$ & Position coordinate of $p^{th}$ grid point\\
$r_s$ & Spread radius\\
$r_c$ & Cutoff radius\\
$v_i$ & Velocity of $i^{th}$ atom\\
$w$ & Weight function\\
$\rho$& Density\\
$\lambda$& Dummy integration variable\\
$\delta$ & Dirac Delta function\\
$\epsilon$& Lennard Jones energy well depth\\
$\sigma$& Lennard Jones zero energy distance\\
\end{tabular}

\textbf{Acknowledgment}. We acknowledge Prof. Xiantao Li of Penn State University for sharing computer code snippets and the helpful discussions with Dr. Maroo.
%
% ...

\bibliography{references}


\end{document}